\documentclass[aps,prb,twocolumn,superscriptaddress,floatfix,longbibliography]{revtex4}
\usepackage{amsfonts}
\usepackage{amssymb}
\usepackage{graphicx}
\usepackage{dcolumn}
\usepackage{bm}
\usepackage{amsmath}
\usepackage[colorlinks,linkcolor=magenta,citecolor=blue,urlcolor=blue]{hyperref}
\usepackage{changes}

\begin{document}

\title{Duality between two generalized Aubry-Andr\'{e} models with exact mobility edges}
\author{Yucheng Wang}
\thanks{These authors contribute equally to this work.\\ Corresponding author: wangyc3@sustech.edu.cn (Yucheng Wang).}
\affiliation{Shenzhen Institute for Quantum Science and Engineering, and Department of Physics,
Southern University of Science and Technology, Shenzhen 518055, China}
\affiliation{International Center for Quantum Materials, School of Physics, Peking University, Beijing 100871, China}
\affiliation{Collaborative Innovation Center of Quantum Matter, Beijing 100871, China}
%\affiliation{$\dagger$ Corresponding author: wangyc3@sustech.edu.cn}
\author{Xu Xia}
%\thanks{These authors contribute equally to this work.}
\thanks{These authors contribute equally to this work.\\ Corresponding author: wangyc3@sustech.edu.cn (Yucheng Wang).}
\affiliation{Chern Institute of Mathematics and LPMC, Nankai University, Tianjin 300071, China}
\author{Yongjian Wang}
%\thanks{These authors contribute equally to this work.}
\thanks{These authors contribute equally to this work.\\ Corresponding author: wangyc3@sustech.edu.cn (Yucheng Wang).}
\affiliation{Academy of Mathematics and Systems Science, Chinese Academy of Sciences, Beijing 100190, China}
\affiliation{University of Chinese Academy of Sciences, Beijing 100049 , China}
\author{Zuohuan Zheng}
\affiliation{Academy of Mathematics and Systems Science, Chinese Academy of Sciences, Beijing 100190, China}
\affiliation{University of Chinese Academy of Sciences, Beijing 100049 , China}
\affiliation{College of Mathematics and Statistics, Hainan Normal University, Haikou, Hainan 571158, China}
\author{Xiong-Jun Liu}
%\thanks{Corresponding author: xiongjunliu@pku.edu.cn}
\affiliation{International Center for Quantum Materials, School of Physics, Peking University, Beijing 100871, China}
\affiliation{Collaborative Innovation Center of Quantum Matter, Beijing 100871, China}
%\affiliation{Beijing Academy of Quantum Information Science, Xibeiwang East Rd, Beijing 100193, China}
\affiliation{CAS Center for Excellence in Topological Quantum Computation, University of Chinese Academy of Sciences, Beijing 100190, China}
\affiliation{Shenzhen Institute for Quantum Science and Engineering, and Department of Physics,
Southern University of Science and Technology, Shenzhen 518055, China}
%\institute{
%\Letter Yucheng Wang \\
%\email{wangyc3@sustech.edu.cn}}

\begin{abstract}
A mobility edge (ME) in energy separating extended from localized states is a central
concept in understanding various fundamental phenomena like the metal-insulator transition in disordered systems. In one-dimensional quasiperiodic systems, there exist a few models with exact MEs, and these models are beneficial to provide exact understanding of ME physics. Here we investigate two widely studied models including exact MEs, one with an exponential hopping and one with a special form of incommensurate on-site potential. We analytically prove that the two models are mutually dual, and further give the numerical verification by calculating the inverse participation ratio and Husimi function. The exact MEs of the two models are also obtained by calculating the localization lengths and using the duality relations. Our result may provide insight into realizing and observing exact MEs in both theory and experiment.
\end{abstract}

%\pacs{74.20.-z, 74.78.-w, 05.30.Rt, 71.10.Pm}
%74.20.-z:theories and models of superconducting state; 74.78.-w:superconduting films and low-dimensional structures;
%05.30.Rt: quantum phase transitions; 71.10.Pm: fermions in reduced dimensions .
\maketitle
%%%%%%%%%%%%%%%%%%%%%%%%
\section{Introduction}
%%%%%%%%%%%%%%%%%%%%%%%%%
Anderson localization (AL)~\cite{Anderson1958}, a fundamental quantum phenomenon in nature, reveals that the single-particle states can become localized due to disorder effect. The quantum phase transition from extended (metal) phase to localized (insulator) phase can occur by increasing the disorder strength in three dimensional (3D) systems. Near the transition point, the mobility edges (MEs) can occur and separate the extended and localized states~\cite{Lagendijk2009,Evers2008}.
ME lie at the heart in understanding various fundamental phenomena such as the metal-insulator transition induced by varying disorder strength or particle number density.
Moreover, a system with ME has strong thermoelectric response~\cite{Whitney2014,Goold2020,Kaoru2017}, and can be applied to thermoelectric devices.
MEs exist widely in 3D systems with random disorder, but for one and two dimensions, the scaling theory~\cite{Anderson1979} shows that all states are localized for arbitrarily small disorder strengths, so no MEs exist.

Unlike random disorder, the quasiperiodic potential can induce the extended-AL transition at a finite strength of the potential even in the 1D systems, which bring about rich interesting physics, e.g., the existence of MEs even in 1D systems~\cite{Biddle,Ganeshan2015,YuWang2020} and non-ergodic critical phases~\cite{YuchengC1,YuchengC2,BoYan2011}.
The most celebrated example with 1D quasiperiodic potential is the Aubry-Andr\'{e} (AA) model~\cite{AA}, described by
$t(\psi_{j+1}+\psi_{j-1})+V\cos(2\pi\beta j+\delta)\psi_{j}=E\psi_{j}$,
where $\psi_j, t, V, \delta$ denote the wavefunction amplitude at site $j$, the nearest-neighbor hopping strength, the strength of quasiperiodic potential, and the phase parameter, respectively, and $\beta$ is an irrational number. The model exhibits a self duality for the transformation between real and momentum spaces at $V=2t$, leading to the extended-localization transition with all the eigenstates of the model being extended (localized) for $V < 2t$ ($V > 2t$). Thus no ME exists for the AA model. This model has been realized in ultracold atomic gases trapped in incommensurate optical lattices, and the localization transition has been observed~\cite{Roati2008}. The existence of the many body localization phase in the quasiperiodic AA model in the presence of weak interactions has also been well established in both theory~\cite{DAHuse2013,YuWang,Das1902} and experiment~\cite{Bloch1,Bloch3}.

By introducing the short-range~\cite{Biddle} or long-range term~\cite{Santos2019}, or breaking the self duality of the AA model~\cite{Xie1988,Biddle2009,Li2017,Ganeshan2015,Zhou2013,Kohmoto2008,Yao2019,YuWang2020}, one can obtain the MEs in the system. However, very few of them can provide the accurate expression of MEs~\cite{Biddle,Xie1988,Ganeshan2015,YuWang2020}, and undoubtedly, these models with exact MEs are beneficial to provide exact understanding of the ME physics for both the non-interacting and interacting systems. In this work, we will focus on the two most commonly used models with exact MEs. One is~\cite{Biddle}
\begin{equation}
 E_1a_n=\sum_{n'\neq n}t_1e^{-p|n-n'|}a_{n'}+V\cos(2\pi\beta n+\delta)a_{n},
\label{ME1}
\end{equation}
where $p>0$, $t_1e^{-p|n-n'|}$ is the hopping rate between the sites $n$ and $n'$ and $V$ is the strength of the quasiperiodic potential. There exists an exponential rather than nearest-neighbor hopping, and this model will become the AA model in the limit $p\rightarrow\infty$. The exact expression of ME is
\begin{equation}
 E_{1c}=V\cosh(p)-t_1.
\label{MEs1}
\end{equation}
The other widely studied model with exact MEs we considered is~\cite{Ganeshan2015}
\begin{equation}
 E_2b_n= t_2(b_{n-1}+b_{n+1})+2\lambda\frac{\cos(2\pi\beta n+\delta)}{1-\alpha\cos(2\pi\beta n+\delta)}b_{n},
\label{ME2}
\end{equation}
where $t_2$ represents the hopping strength between neighboring sites, $\lambda$ and $\alpha$ ($\alpha\in (-1,1)$) represent the on site modulation strength and the deformation parameter, respectively, and when $\alpha=0$, this model reduces to the AA model.
The exact expression of the ME is
\begin{equation}
 E_{2c}=2sgn(\lambda)(|t_2|-|\lambda|)/\alpha.
\label{MEs2}
\end{equation}
For convenience, we call the above-mentioned first (second) model Model I (II).
The two models have widely been used to study the ME physics, e.g., the dynamical behavior of a system with MEs~\cite{Xu2020}, fate of MEs in the presence of interactions~\cite{LiX2015,Modak2015,LiX2016,Gao2019} or non-Hermitian term~\cite{Chen2020,Yong2020,LiuT2020}.

In recent years, MEs have been observed in disordered systems~\cite{Bouyer2012,McGehee2013,Modugno2015} and quasiperiodic systems~\cite{BlochME,Gadway2018,BlochME2,Gadway2020} in experiments based on ultracold atoms. In particular, the recent work~\cite{Gadway2020}
has accurately realized the Model II (\ref{ME2}) by using synthetic lattices of laser-coupled atomic momentum modes, and accurately detected the location of MEs in the absence and presence of interactions.

This study is motivated by two nontrivial questions raised here. First, is there any profound relation between the above two models even they seem to be quite different, and whether the Model I can be accurately realized in experiment? Secondly, can the localization lengths of the states in two models can be exactly computed, which clearly necessitates to go beyond the dual transformation applied to determine the ME in the previous studies. %The localization length is still not clear, and whether the Model I can be accurately realized and the location of MEs can be accurately detected.
Answering these questions is important to unveil the fundamental properties of the two important models.
In this work, we prove that the above two generalized AA models (Eq.~(\ref{ME1}) and Eq.~(\ref{ME2})) have the mutually dual relation, and further provide exact study of the localization properties of the states.
In particular, the Hamiltonian of the Model II can be written as a
tri-diagonal matrix, whose ME expression can be obtained by using a self-consistent theory~\cite{Duthie} or by calculating the
localization length numerically by using the recursive methods~\cite{Yizhang} and analytically by using Avila's global theory~\cite{YuWang2020,Avila,WangYJ}. By using the duality relation between Model I and Model II, we further determine the exact ME of Model I, whose localization length is difficult to be directly computed and the self-consistent theory can not also be used due to the exponential hopping. With the dual relation proved in this work, the recent experimental work~\cite{Gadway2020} that realized the Model II (\ref{ME2}) in momentum space could be regarded to have also effectively realized the Model I in real space.

%%%%%%%%%%%%%%%%%%%%%%%%%%%%%%%%%%%%%%%%
\section{analytical and numerical results}
%\label{non-interacting}
%%%%%%%%%%%%%%%%%%%%%%%%%%%%%%%%%%%%%%%
%%%%%%%%%%%%%%%%%%%%%%%%%%%%%%%%%%%%%%%%
\subsection{analytical derivation for dual relations}
%%%%%%%%%%%%%%%%%%%%%%%%%%%%%%%%%%%%%%%
We firstly analytically establish the duality between the Model I and Model II. Since the phase offset $\delta$ is redundant in the context of localization~\cite{Ganeshan2015}, without loss of generality, we set $\delta=0$.
We start from the Model I (\ref{ME1}), and introduce the transformation
\begin{equation}
 a_j=\frac{1}{\sqrt{L}}\sum_mb_me^{-i2\pi m\beta j},
\label{trans}
\end{equation}
where
\begin{equation}
 b_m=\frac{1}{\sqrt{L}}\sum_ja_je^{i2\pi m\beta j},
\label{trans2}
\end{equation}
By using the transformation (\ref{trans}), Eq.~(\ref{ME1}) becomes
\begin{widetext}
\begin{eqnarray}
E_1\frac{1}{\sqrt{L}}\sum_m b_me^{-i2\pi m\beta n}=\sum_{n'\neq n}t_1e^{-p|n-n'|}\frac{1}{\sqrt{L}}\sum_m b_me^{-i2\pi m\beta n'}+V\cos(2\pi\beta n)\frac{1}{\sqrt{L}}\sum_m b_me^{-i2\pi m\beta n}.
\label{Lam1}
\end{eqnarray}
\end{widetext}
Then we rewrite the first term on the right side of the equation $\sum_{n'\neq n}t_1e^{-p|n-n'|}\frac{1}{\sqrt{L}}\sum_m b_me^{-i2\pi m\beta n'}$ as
%\begin{widetext}
%\begin{eqnarray}
%\sum_{n'\neq n}t_1e^{-p|n-n'|}\frac{1}{\sqrt{L}}\sum_m b_me^{-i2\pi m\beta n'} \nonumber\\
%=\frac{1}{\sqrt{L}}\sum_m b_me^{-i2\pi m\beta n}\sum_{n'\neq n}t_1e^{-p|n-n'|}e^{-i2\pi m\beta (n'-n)}.
%\label{Lam2}
%\end{eqnarray}
%\end{widetext}
$\frac{1}{\sqrt{L}}\sum_m b_me^{-i2\pi m\beta n}\sum_{n'\neq n}t_1e^{-p|n-n'|}e^{-i2\pi m\beta (n'-n)}$, where
$\sum_{n'\neq n}t_1e^{-p|n-n'|}e^{-i2\pi m\beta (n'-n)}$ is the summation of a geometric sequence, and one can obtain that it equals to
$\frac{2t_1(-e^{-2p}+e^{-p}\cos(2\pi m\beta))}{1+e^{-2p}-2e^{-p}\cos(2\pi m\beta)}$.
Then Eq.~(\ref{Lam1}) can be written as
\begin{widetext}
\begin{eqnarray}
E_1\frac{1}{\sqrt{L}}\sum_m b_me^{-i2\pi m\beta n}
      = \frac{1}{\sqrt{L}}\sum_m b_me^{-i2\pi m\beta n}\frac{2t_1(-e^{-2p}+e^{-p}\cos(2\pi m\beta))}{1+e^{-2p}-2e^{-p}\cos(2\pi m\beta)} %\nonumber\\
      + \frac{V}{2}\frac{1}{\sqrt{L}}\sum_m (b_{m-1}+b_{m+1})e^{-i2\pi m\beta n},
\notag
\end{eqnarray}
\end{widetext}
Utilizing the above formula, one can directly obtain
%\begin{widetext}
\begin{eqnarray}
E_1 b_m=\frac{2t_1(-e^{-2p}+e^{-p}\cos(2\pi m\beta))}{1+e^{-2p}-2e^{-p}\cos(2\pi m\beta)}b_m \nonumber\\
+\frac{V}{2}(b_{m-1}+b_{m+1}).\qquad \qquad \qquad
\label{Lambda6}
\end{eqnarray}
%\end{widetext}
Let
\begin{subequations}\label{Lamba7X}
\begin{eqnarray}
t_2 = \frac{V}{2},\qquad \qquad \qquad \quad
%\label{Lambda7a}
\alpha = \frac{2e^{-p}}{1+e^{-2p}},\qquad \qquad \quad
\label{Lambda7a}\\
E_2 = E_1+\frac{2t_1e^{-2p}}{1+e^{-2p}},\qquad
%\label{Lambda7c}
\lambda = \frac{t_1(-e^{-3p}+e^{-p})}{(1+e^{-2p})^2},\qquad
\label{Lambda7b}
\end{eqnarray}
\end{subequations}
then Eq.~(\ref{Lambda6}) is equivalent to Eq.~(\ref{ME2}). Therefore, Model I (Eq.~(\ref{ME1})) and Model II (Eq.~(\ref{ME2})) are mutually dual.

%%%%%%%%%%%%%%%%%%%%%%%%%%%%%%%%%%%%%%%%
\subsection{Localization lengths and mobility edges}
%%%%%%%%%%%%%%%%%%%%%%%%%%%%%%%%%%%%%%%
Since the two models (Model I and Model II) are mutually dual, we can obtain some properties of one model from the other model.
Model I is not exactly solvable due to the existence of the exponential hopping, but
Model II can be written as a tri-diagonal matrix, whose all states' extended and localized properties can be analytically obtained by using the Avila's
global theory~\cite{YuWang2020,Avila,WangYJ}. We firstly represent the Eq.~(\ref{ME2}) in the transfer matrix form,
\begin{equation}
\left(
\begin{array}{c}
b_{n+1} \\
b_{n}
\end{array}
\right)=T^n\left(
\begin{array}{c}
b_{n} \\
b_{n-1}
\end{array}
\right)
\notag
\end{equation}
where the transfer matrix $T^{n}$ is given by
\begin{equation}
T^{n}=\left(
\begin{array}{cc}
\frac{E_2}{t_2}-\frac{2\lambda}{t_2}\frac{\cos(2\pi\beta n+\delta)}{1-\alpha\cos(2\pi\beta n+\delta)} & -1 \\
1 & 0
\end{array}
\right)
\label{TM}
\end{equation}
Using the transfer matrix, one can define and compute the Lyapunov exponent (LE),$$\gamma(E)=\lim_{n\rightarrow \infty}\frac{1}{2\pi L} \int \ln  \|T_L(\delta)\| d\delta,$$ where  $T_L=\prod_{n=1}^{L}T^{n}$ and $\|T_L\|$ denotes the norm of the matrix $T_L$. The LE can be exactly obtained by using Avila's global
theory~\cite{Avila,WangYJ}, and the details for the calculation are put in the Appendix. By the LE, we can obtain the localization length $\xi$, which is the reciprocal of the LE, i.e.,
\begin{equation}
\xi(E)=\frac{1}{\gamma(E)}=\frac{1}{\ln|\frac{|\alpha E+2\lambda|+\sqrt{(\alpha E+2\lambda)^2-4\alpha^2}}{2(t+\sqrt{t^2-\alpha^2})}|}.
\label{LL}
\end{equation}
When $|\frac{|\alpha E+2\lambda|+\sqrt{(\alpha E+2\lambda)^2-4\alpha^2}}{2(t+\sqrt{t^2-\alpha^2})}|>1 (=1)$, $\xi$ is a finite (infinite) value and the corresponding state is localized (decolized). Thus the critical points and MEs are determined
by $|\frac{|\alpha E_{2c}+2\lambda|+\sqrt{(\alpha E_{2c}+2\lambda)^2-4\alpha^2}}{2(t+\sqrt{t^2-\alpha^2})}|=1$, which can give the ME expression (Eq.~(\ref{MEs2})) (see the Appendix). The ME expression of Model II can also be analytical obtained by using a self-consistent theory~\cite{Duthie}.

Naturally, combining Eqs.~(\eqref{Lambda7a},\eqref{Lambda7b}) and the expression Eq.~(\ref{MEs2}) of Model II's ME, one can obtain the expression Eq.~(\ref{MEs1}) of Model I's ME.

%%%%%%%%%%%%%%%%%%%%%%%%%%%%%%%%%%%%%%%%%%%%%%%
\begin{figure}[t]
\hspace*{-0.3cm}
\includegraphics[width=0.51\textwidth]{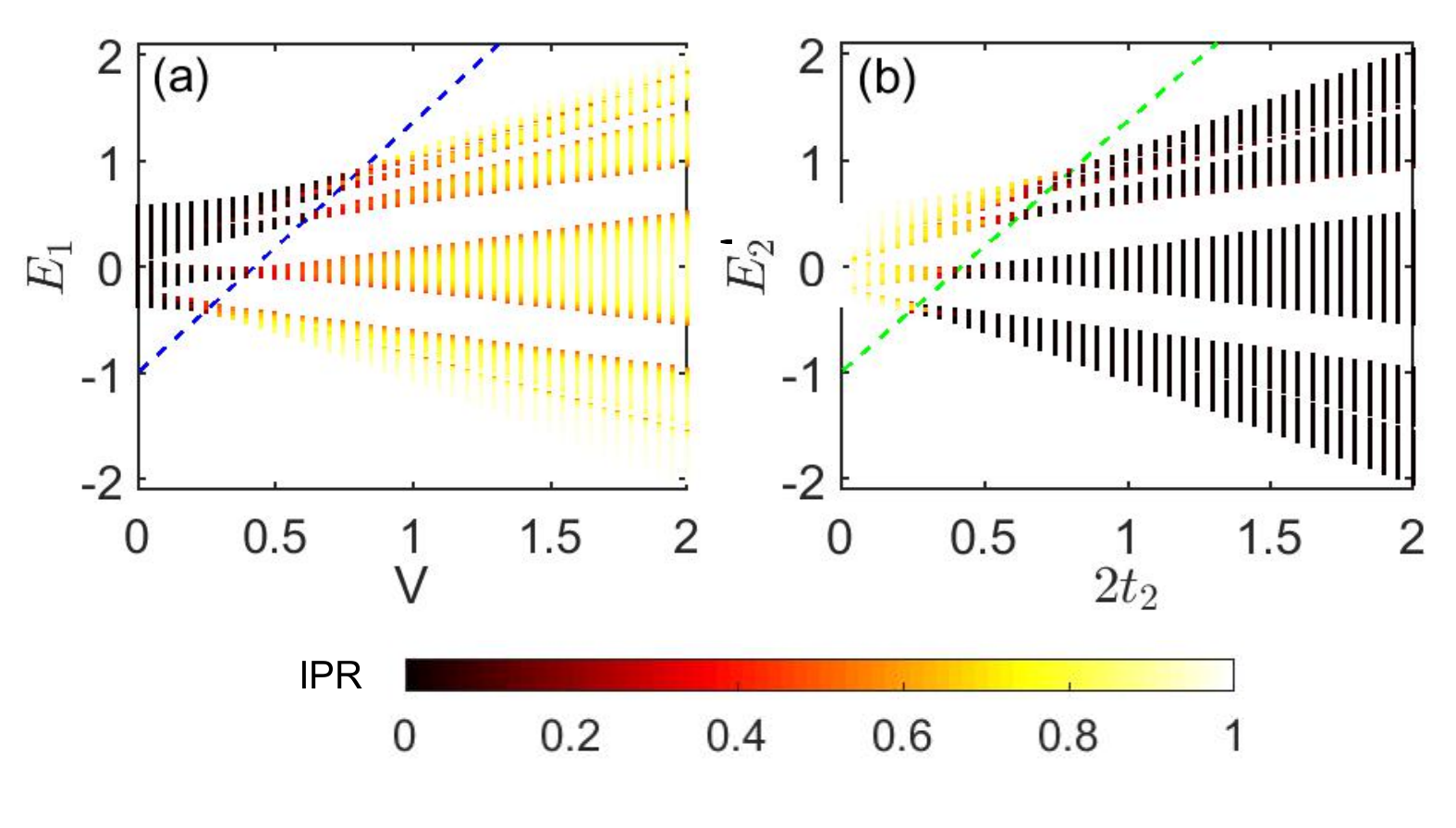}
\caption{\label{01}
 (a) IPR of different eigenstates as a function of the corresponding eigenvalues $E_1$ and quasiperiodic potential strength $V$ with fixed $p=1.5$ and $t_1=1$ in Model I, (b) IPR as a function of $E'_2$ and $2t_2$ with fixed $\alpha=0.4251$ and $\lambda=0.1924$ in Model II. Here the values of $\alpha$ and $\lambda$ are obtained from Eq.~(\ref{Lambda7a}) and Eq.~(\ref{Lambda7b}) with fixed $p=1.5$ and $t_1=1$. $E'_2=E_2-\frac{2t_1e^{-2p}}{1+e^{-2p}}$. The blue and green dotted line in (a) and (b) are obtained from Eq.~(\ref{MEs1}). Here we fix $\beta=(\sqrt{5}-1)/2$ and the size $L=500$.}
\end{figure}
%%%%%%%%%%%%%%%%%%%%%%%%%%%%%%%%%%%%%%%%%%%%%%%%

%%%%%%%%%%%%%%%%%%%%%%%%%%%%%%%%%%%%%%%%
\subsection{numerical results}
%%%%%%%%%%%%%%%%%%%%%%%%%%%%%%%%%%%%%%%
Now we display the numerical evidence for the dual relation. The numerical results are obtained by calculating the inverse participation ratio (IPR)~\cite{Evers2008} $IPR(\kappa)=\sum_{j=1}^{L}|\psi_{\kappa,j}|^4$, where $\psi_{\kappa}$ is the $\kappa$-th eigenstate. It is known that tends to zero in the thermodynamic limit for extended states, but approaches to a finite value of $O(1)$ for a localized state. Fig.~\ref{01} (a) shows the energy eigenvalues and the IPR of the corresponding eigenstates for Model I as a function of $V$ under open boundary conditions. The dotted line represents the ME given in Eq.~(\ref{MEs1}). We see that IPR values are approximately zero for energies above the ME and are finite for energies below the ME. In fig.~\ref{01} (a), we take $p=1.5$ and $t_1=1$, which can give $\alpha=0.4251$ and $\lambda=0.1924$ from Eq.~(\ref{Lambda7a}) and Eq.~(\ref{Lambda7b}). Then fixing $t_2$, we can diagonalize the Model II and obtain its eigenvalues $E_2$ and the IPR of the corresponding eigenstates. Fig.~\ref{01} (b) shows the IPR as a function of $E'_2$ and $2t_2$, where $E'_2=E_2-\frac{2t_1e^{-2p}}{1+e^{-2p}}$. Due to $V=2t_2$, we take the horizontal axis being $2t_2$ to compare with fig.~\ref{01} (a). One can see that the two energy spectrum in fig.~\ref{01} (a) and fig.~\ref{01} (b) are exactly the same, but the IPR values in fig.~\ref{01} (b) are finite for energies above the ME and are approximately zero for energies below the ME, which is contrary to fig.~\ref{01} (a), indicating that the Model I and Model II are mutually dual.

\begin{figure}[t]
\hspace*{-0.3cm}
\includegraphics[width=0.51\textwidth]{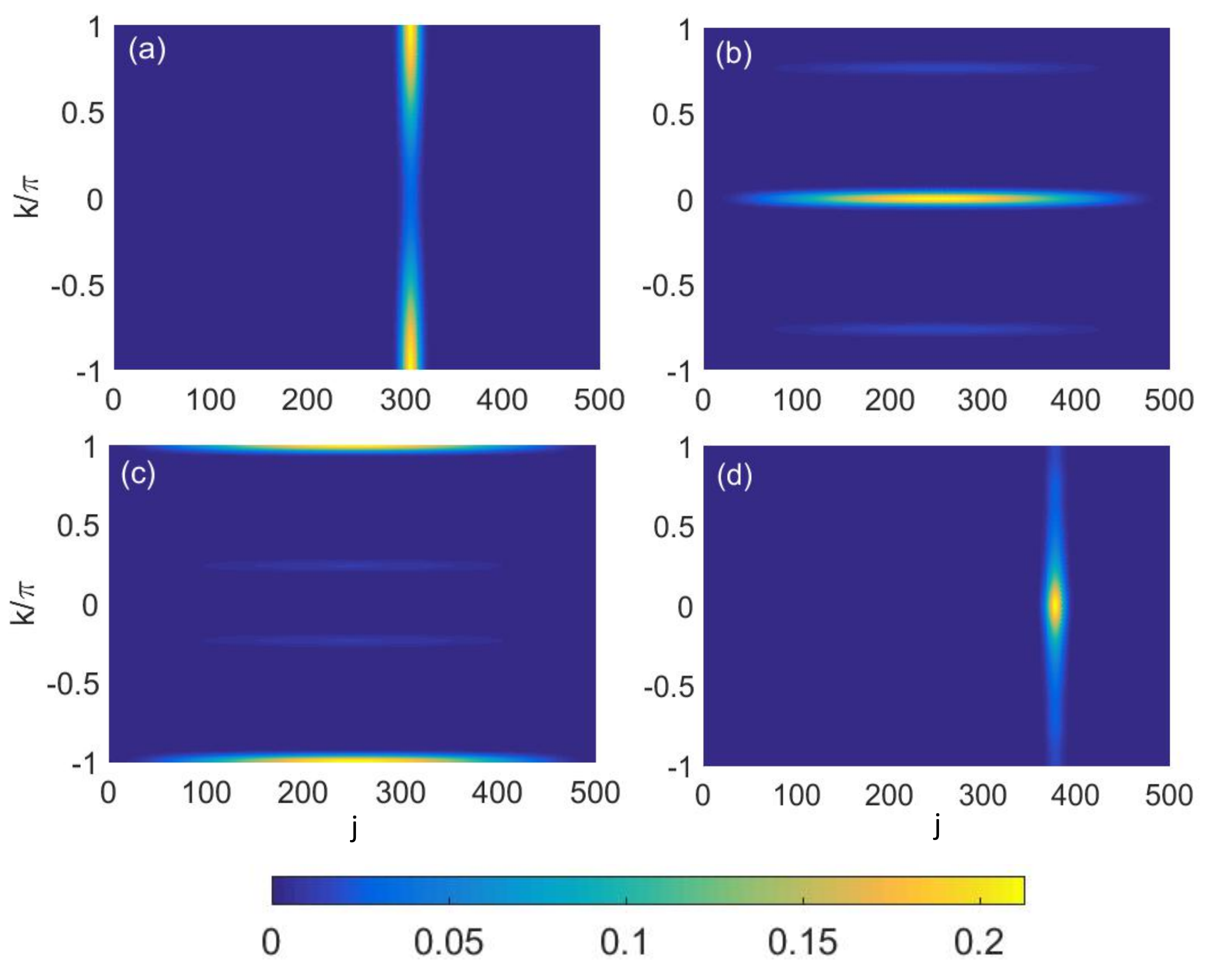}
\caption{\label{02}
 The Husimi function $\rho(j,k)$ for the eigenstate corresponding the lowest energy in (a) and (c) and highest energy in (b) and (d). (a) and (b) correspond to Model I with $p=1.5$, $t_1=1$ and $V=0.5$, (c) and (d) correspond to Model II with $\alpha=0.4251$, $\lambda=0.1924$ and $t_2=0.25$. Here we fix $\beta=(\sqrt{5}-1)/2$ and the size $L=500$.}
\end{figure}
%%%%%%%%%%%%%%%%%%%%%%%%%%%%%%%%%%%%%%%%%%%%%%%%

Besides the IPR, we further introduce the Husimi function~\cite{Husimi,Harriman,Varga2004,Feist2006} to gain a better intuition for the localization behavior in both the real space and momentum space. It is given by
\begin{equation}
 \rho(j_0,k_0)=|\langle j_0,k_0|\psi\rangle|^2.
\label{HF}
\end{equation}
%%%%%%%%%%%%%%%%%%%%%%%%%%%%%%%%%%%%%%%%%%%%%%%
Here the Husimi function is the probability density function for finding the system with state $|\psi\rangle$ in a minimum-uncertainty state centered at $j_0$ in coordinate space and at $k_0$ in momentum space. Note that while the momentum
is not a good quantum number here, the projection on the selected $k_0$ can be done. Using the minimal uncertainty state in real space~\cite{Varga2004}
%\begin{widetext}
%\begin{eqnarray}
\begin{equation}
\langle j|j_0, k_0\rangle=(\frac{1}{2\pi\sigma^2})^{1/4}exp(-\frac{(j-j_0)^2}{4\sigma^2}+ik_0(j+j_0/2)),
\notag
\end{equation}
%\end{eqnarray}
%\end{widetext}
where $\sigma$ is taken as $\sigma=\sqrt{\frac{L}{4\pi}}$, and inserting $\sum_j|j\rangle\langle j|$ in Eq.~(\ref{HF}), one can obtain the Husimi function, as shown in fig.~\ref{02}. From fig.~\ref{02} (a), we see that there exists one vertical stripe for the lowest state of Model I indicating that the considered state is localized in real space and extended in momentum space.
By contrast, fig.~\ref{02} (b) shows three horizontal stripes symmetrically placed with respect to $k=0$ for the highest state of Model I indicating that the state is localized in momentum space and extended in real space. Comparing fig.~\ref{02} (a) with fig.~\ref{02} (b), we see that there exists a ME for the Model I with $p=1.5, t_1=1$ and $V=0.5$, and the eigenstates are spatially localized and extended below and above the ME. The same analysis applies to the Model II (see fig.~\ref{02} (c) and (d)), whose eigenstates are spatially extended and localized below and above the ME, and it is consistent with our conclusion that Model I and Model II are mutually dual.

%%%%%%%%%%%%%%%%%%%%%%%%%%%%%%%%%%%%%%%%
\section{summary and discussion}
%\label{non-interacting}
%%%%%%%%%%%%%%%%%%%%%%%%%%%%%%%%%%%%%%%
We have analytically proven that the two widely studied models (Model I (\ref{ME1}) and Model II (\ref{ME2})) with exact MEs are mutually dual. By using the
Avila's global theory, one can give the localization length and ME expressions of Model II. Then Model I's ME expressions can be obtained by using the dual relation. We further numerically verified our result by calculating the IPR and Husimi function. Our conclusion will make the ME's study more convenient. In theory, studying the physical properties of one of the models, one can deduce the corresponding properties of the other model. On the other hand, we provide the new approach to study the system with the exponential hopping, whose some properties are difficult to investigated both numerically and analytically but can be obtained for its dual model. In experiment, the realization of the Model I and the detection of the location of ME can be replaced by detecting the location of Model II's ME in momentum space, which has been realized and detected recently~\cite{Gadway2020}.
%When adding interactions, the Model I have been widely studied theoretically~\cite{LiX2015,LiX2016,Gao2019}, and the theoretical results may be verified by studying the Model II experimentally~\cite{Gadway2020}.

\begin{acknowledgments}
Yucheng Wang and X.-J. Liu are supported by National Nature Science Foundation of China (11825401, 11761161003, and 11921005), the National Key R\&D Program of China (2016YFA0301604), Guangdong Innovative and Entrepreneurial Research Team Program (No.2016ZT06D348), the Science, Technology and Innovation Commission of Shenzhen Municipality (KYTDPT20181011104202253), and the Strategic Priority Research Program of Chinese Academy of Science (Grant No. XDB28000000). X. Xia is supported by NanKai Zhide Foundation. Yongjian Wang is supported by the National Natural Science Foudation of China (No. 12061031).Zuohuan Zheng acknowledges financial supports of the NSF of China (No. 12031020, 11671382), CAS Key
Project of Frontier Sciences (No. QYZDJ-SSW-JSC003), the Key Lab. of Random Complex Structures and Data
Sciences CAS and National Center for Mathematics and
Interdisciplinary Sciences CAS.
\end{acknowledgments}
%This will simplify the ME's research in both theory and experiment. For example, in experiment, it is difficult to accurately realize the Model I (\ref{ME1}) and detect the location of ME, but owing to the dual relation, the above-mentioned work~\cite{Gadway2020} that realized the Model II (\ref{ME2}) in momentum space can be considered to have also realized the Model I in real space and detected the location of ME.
%\clearpage

%%%%%%%%%%%%%%%%%%%%%%
\appendix
%%%%%%%%%%%%%%%%%%%%%%%%%%%%
\section{Details for localization length}
%%%%%%%%%%%%%%%%%%%%%%%%%%%%%%%
In this appendix, we give the detail of the derivation of Lyapunov exponents. For convenience, we set the hopping strength $t_2=1$. The transfer matrix (\ref{TM}) can be decomposed into two parts, $T^{n}=A^{n}B^{n}$, where
\begin{equation}
 A^n=\frac{1}{1-\alpha\cos(2\pi\beta n+\delta)},
\label{An}
\end{equation}
and
\begin{equation}
B^{n}=\left(
\begin{array}{cc}
B_{11} & B_{12} \\
B_{21} & 0
\end{array}
\right)
\label{Bn}
\end{equation}
with $B_{11}=E(1-\alpha\cos(2\pi\beta n+\delta))-2\lambda\cos(2\pi\beta n+\delta)$ and $B_{21}=-B_{12}=1-\alpha\cos(2\pi\beta n+\delta)$.
Then
\begin{equation}
\gamma(E)=\gamma_A(E)+\gamma_{B}(E),
\label{LEZ}
\end{equation}
where $\gamma_A=\lim_{n\rightarrow \infty}\frac{1}{2\pi L} \int \ln  \|A_L(\delta)\| d\delta$ with $A_L=\prod_{n=1}^{L}A^{n}$. By the ergodic theory, $\gamma_A(E)=\frac{1}{2\pi}\int^{2\pi}_0 \ln(\frac{1}{1-\alpha\cos(\delta)})d\delta=-\ln|\frac{1+\sqrt{1-\alpha^2}}{2}|$~\cite{Gradshteyn,Longhi}.
In Eq.~(\ref{LEZ}), $\gamma_B=\lim_{n\rightarrow \infty}\frac{1}{2\pi L} \int \ln  \|B_L(\delta)\| d\delta$ with $B_L=\prod_{n=1}^{L}B^{n}$.
Below we calculate $\gamma_B$ relies on Avila's global theory~\cite{Avila}. We firstly complexify the phase, i.e., $B_{11}=E(1-\alpha\cos(2\pi\beta n+\delta+i\epsilon))-2\lambda\cos(2\pi\beta n+\delta+i\epsilon)$, $B_{21}=-B_{12}=1-\alpha\cos(2\pi\beta n+\delta+i\epsilon)$. Then let $\epsilon$ tends to infinity, the matrix $B^{n}$ becomes
\begin{widetext}
\begin{equation}
B^{n}(\delta+i\epsilon)=\frac{e^{2\pi\epsilon}e^{i(2\pi\beta n+\delta)}}{2}\left(
\begin{array}{cc}
-\alpha E-2\lambda & \alpha \\
-\alpha & 0
\end{array}
\right)
+o(1)
\end{equation}
\end{widetext}
Thus we have $\gamma_B(E,\epsilon)=2\pi\epsilon+\ln|\frac{|\alpha E+2\lambda|+\sqrt{(\alpha E+2\lambda)^2-4\alpha^2}}{4}|+o(1)$. By the global theory~\cite{Avila}, we obtain $\gamma_B(E)=\ln|\frac{|\alpha E+2\lambda|+\sqrt{(\alpha E+2\lambda)^2-4\alpha^2}}{4}|$. Plugging $\gamma_A(E)$ and $\gamma_B(E)$ into Eq.~(\ref{LEZ}), we have $\gamma(E)=\ln|\frac{|\alpha E+2\lambda|+\sqrt{(\alpha E+2\lambda)^2-4\alpha^2}}{2(1+\sqrt{1-\alpha^2})}|$, which give the localization length, as shown in Eq.~(\ref{LL}). As our discussions in the main text, MEs satisfy
 $|\frac{|\alpha E+2\lambda|+\sqrt{(\alpha E+2\lambda)^2-4\alpha^2}}{2(1+\sqrt{1-\alpha^2})}|=1$. Now we set $P=\alpha E+2\lambda$, then MEs satisfy
 $|P|+\sqrt{P^2-4\alpha^2}=2(1+\sqrt{1-\alpha^2})$, which give $|P|=2$, i.e., $|\alpha E+2\lambda|=2$, which can give Eq.~(\ref{MEs2}) in the main text.

%%%%%%%%%%%%%%%%%%%%%%%%%%%%%%%%%%%%%%%%%%%%%%

\end{document}